\documentclass[apj,twocolumn]{openjournal}

\usepackage{xcolor}
\usepackage[utf8]{inputenc}
\usepackage[english]{babel}

\usepackage{orcidlink}
\usepackage{makecell}
\usepackage{amsmath}
\usepackage{booktabs}
\usepackage{multirow}
\usepackage{placeins}

\usepackage{siunitx}
\DeclareSIUnit{\kms}{\kilo\meter\per\second}
\DeclareSIUnit{\erg}{erg}
\DeclareSIUnit{\ergs}{\erg\per\second}
\DeclareSIUnit{\keV}{\kilo\electronvolt}
\DeclareSIUnit \parsec {pc}
\DeclareSIUnit{\kpc}{\kilo\parsec}
\DeclareSIUnit{\mpc}{\mega\parsec}
\definecolor{linkcolor}{rgb}{0.0,0.3,0.5}
\usepackage{hyperref}
\hypersetup{
    unicode, 
    colorlinks=true,
    linkcolor=linkcolor,
    citecolor=linkcolor,
    filecolor=linkcolor,
    urlcolor=linkcolor,
}
\usepackage{cleveref}

\crefname{figure}{Figure}{Figures}
\crefname{equation}{Equation}{Equations}
\crefname{table}{Table}{Tables}
\crefname{section}{Section}{Sections}
\crefname{subsection}{Subsection}{Subsections}

\crefformat{figure}{Figure #2#1#3}
\crefformat{table}{Table #2#1#3}
\crefformat{equation}{Equation #2#1#3}
\crefformat{section}{Section #2#1#3}
\crefformat{subsection}{Subsection #2#1#3}

\Crefformat{figure}{Figure #2#1#3}
\Crefformat{table}{Table #2#1#3}
\Crefformat{equation}{Equation #2#1#3}
\Crefformat{section}{Section #2#1#3}
\Crefformat{subsection}{Subsection #2#1#3}

\newcommand{\uat}[2]{\href{https://astrothesaurus.org/uat/#2}{#1\ (#2)}}

\makeatletter

\renewcommand{\@tablenotes}{}

\newcommand{\tablenotesreset}{\gdef\@tablenotes{}}

\renewcommand{\tablenotemark}[1]{\textsuperscript{#1}}

\makeatother



\makeatletter

\@ifundefined{LaTeXML}{%
  \renewcommand{\tablenotetext}[2]{%
  \g@addto@macro\@tablenotes{\makebox[0.5em][l]{\textsuperscript{\vphantom{b}#1}}#2\newline}%
}

\newcommand{\printtablenotes}[1]{%
  \noalign{\vskip 2pt}%
  \multicolumn{#1}{@{}l@{}}{%
    \parbox[t]{\linewidth}{%
      \footnotesize\raggedright\setlength{\parindent}{0pt}%
      \@tablenotes
    }%
  }\\%
}
}{%
  \renewcommand{\arcmin}{\ensuremath{^\prime}}
  \renewcommand{\arcsec}{\ensuremath{^{\prime\prime}}}

  \renewcommand{\tablenotetext}[2]{%
  \g@addto@macro\@tablenotes{\textsuperscript{#1}#2\newline}%
}

\newcommand{\printtablenotes}[1]{%
  \noalign{\vskip 2pt}%
  \multicolumn{#1}{@{}l@{}}{%
    \begin{minipage}{\linewidth}%
      \footnotesize\raggedright\setlength{\parindent}{0pt}%
      \@tablenotes
    \end{minipage}%
  }\\%
}

}
\makeatother

\makeatletter

\makeatletter
\newcommand{\pdffig}[1]{\@ifundefined{LaTeXML}{#1}{}}
\newcommand{\htmlfig}[1]{\@ifundefined{LaTeXML}{}{#1}}
\makeatother

\begin{document}

\title{Fast Astronomical Transients in Archival Photographic Plates: Using optical 
aberrations as a tool for discerning real images, from plate artifacts}

\makeatletter
\@ifundefined{LaTeXML}{%
  \author{
    {Ivo Busko\orcidlink{0009-0009-6828-1164}}, 
  }
  \email{ivobusko@gmail.com}
  \affiliation{Independent Researcher, Retired Developer at NASA/AURA/STScI}
}{%

  \author{Ivo Busko\orcidlink{0009-0009-6828-1164}}
  \affiliation{Independent Researcher, Retired Developer at NASA/AURA/STScI}
  \email{ivobusko@gmail.com}
}

\makeatother

\begin{abstract}
The detection of fast astronomical transients in photographic plates from the Palomar 
sky surveys conducted in the 1950s, was subject to the criticism that such transients 
could be just the effect of otherwise unaccounted for plate artifacts. 
In this paper, we show that transient images exhibit the coma aberration pattern 
expected from off-axis point sources recorded through the telescope optics, a signature 
that plate artifacts cannot naturally reproduce. Although
the data does not by themselves establish the physical origin of the light that generated
the images, they lend support to hypotheses that do not rely on instrumental effects to
explain transients. 
\end{abstract}


\maketitle

\section{Introduction}\label{sec:Introduction}

In this paper, we present first results from a search of optical transients carried out in 
the context of the VASCO Project (Vanishing \& Appearing Sources during a Century 
of Observations, \citealt{Villarroel2020} ), 
but using different methodology and data. 
Methodology and data were described previously (\citealt{Busko2026b}, henceforth Paper I).  
In short, we resort to data provided by the Archives of Photographic PLates for 
Astronomical USE (APPLAUSE)\footnote{\url{ https://www.plate-archive.org/cms/home/}}, 
and we search for individual transient events, as opposed to the statistical approach 
used by the VASCO team. Using a different data set than the digitized versions of the 
Palomar Observatory Sky Surveys (POSS, \citealt{DPOSS-II}) used by VASCO enables 
us to independently test 
their findings, although at a cost: because the POSS was conducted as a homogeneous 
survey over most of the sky, with a single telescope, the resulting data set can be reliably 
used to build statistically complete samples from which sound statistical information 
can be gathered. The APPLAUSE archive, on the other hand, being a collection of 
photographic plates taken for different scientific purposes, and using different telescopes, 
does not lend itself easily to the construction of large and homogeneous data sets. 
This is the underlying reason why we adopted the search procedure described in Paper 1.

Regardless of the data and method used, the main problem is that a plate artifact
can potentially be mistaken as a legitimate transient. Plate artifacts result from a large number 
of causes. Some are simple to understand, such as dust particles and micro-hairs, deposited
or attached both on the plate surfaces and on the scanner bed. Also, finger prints, scratches, 
and blemishes introduced by careless manipulation. Other causes are more exotic in nature
but no less damaging, such as ambient radiation, 
chemical issues during plate development, manufacture defects, digital scanning artifacts. 
An important issue is aging: plates stored for many decades inside a paper envelope 
can develop markings and spots caused by the envelope slowly leaching chemicals. 
Particles firmly attached, or even embedded in the emulsion layer, can slowly leach 
chemicals into the emulsion as well. Most of these artifacts can be easily filtered out 
after the plate is scanned, with software such as SExtractor \citep{1996A&AS..117..393B}.
But a small but persistent subset will remain, because they mimic so well the appearance 
of real stars that were truly imaged by the telescope. This problem was addressed
by a number of works in the 1990s, especially in relation to the identification attempts
of Gamma-Ray Burst (GRB) optical counterparts in historical collections of photographic plates.
To this day, it remains an unsolved problem: no GRB counterparts were ever unambiguously
identified in the optical by these techniques (see \cite{watters2026} for a comprehensive 
review).

Thus, an often raised criticism to the proposal that transients were recorded by these old 
plates is that they could be just plate artifacts that somehow managed to elude the analysis 
and statistical tools used to find them. Although a number of works already produced
 statistical evidence that true transients exist 
 \citep{Villarroel2020, villaroel2025aligned, 2604.18799}, the subject is still under debate.
 For instance, \cite{watters2026} present several arguments that suggest the existence
 of transients is still not well stablished.
 Thus, it would be reassuring to be able to confirm transient existence through independent data
 sets and methodologies.

\section{Coma optical aberration}

In the project described on Paper 1, we focussed on plates obtained with the 
{\it Hamburger Sternwarte Großer 
Schmidtspiegel} 1.2-m camera. Being of Schmidt design, the images it produces are 
remarkably free of optical aberrations that can deform the image produced by a 
point source. 
However, this much sought-after feature, so useful in most astronomical research,
 turned out to be a liability for this project. The reason is that the sharp and very round
 images the camera creates from point sources, are difficult to tell apart from some types of artifacts. 
 There is a small but significant population of artifacts that look very round, and sometimes even 
 a bit sharper that star images. To the point that one cannot distinguish between them 
 based on common measures such as radial profiles, widths, circularity figures-of-merit,
 and others. 
 
 However, the APPLAUSE archive hosts data from many historical telescopes, and some
 of those are prone, sometimes very significantly, to optical aberrations. A common type 
 of aberration in simple telescopes is {\it coma} 
 (from the latin {\it coma} or greek {\it koma} meaning ``hair of a comet''). 
 It is caused by light rays coming into the telescope at an angle with 
 the optical axis, and thus losing their symmetry in relation to the optical surfaces. This 
 results in elongated images that resemble a tiny comet with its tail \citep{coma}.  
 
\begin{figure*}[htb!]
    \centering
    \begin{minipage}{0.49\linewidth}
        \includegraphics[scale=0.3]{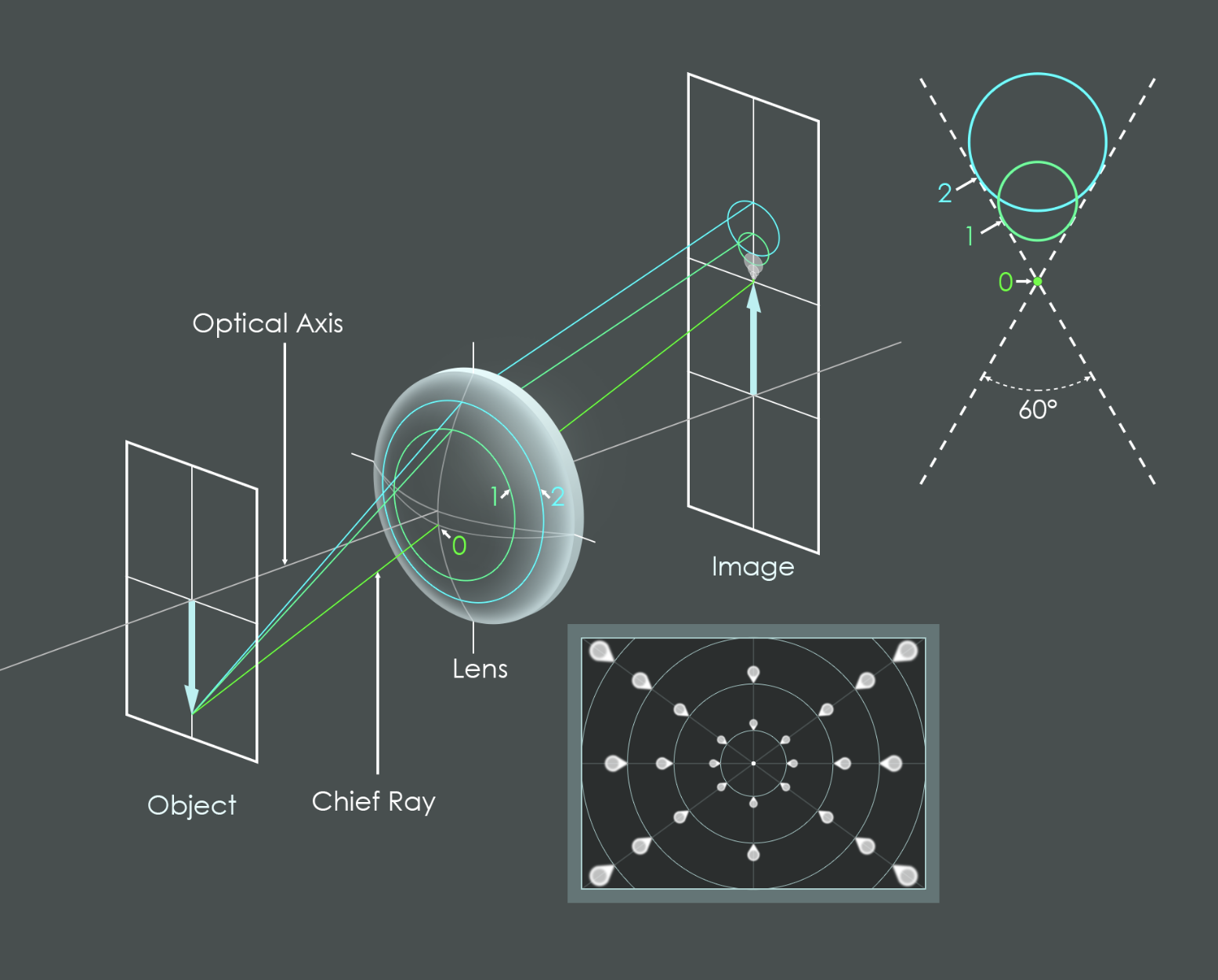}
    \end{minipage}
    \hfill
    \begin{minipage}{0.49\linewidth}
        \includegraphics[scale=0.67]{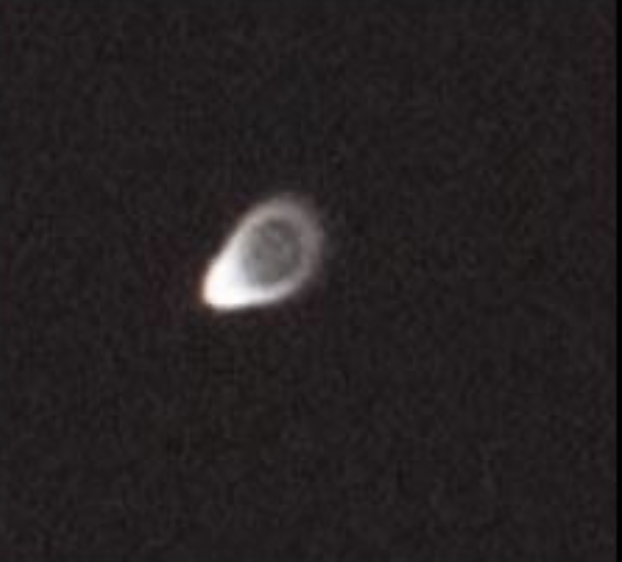}
    \end{minipage}
    \caption{{\bf Coma:} light rays from a point source, coming at an angle into the optics, 
    result in an elongated light distribution at the focal surface. The light distribution 
    is built from continuously enlarged circles, as one moves away from the chief ray
    towards the plate edge.  
    These circles are essentially defocussed images of the point source, that is, they
    are images of the lens {\it pupil}.
    The locus where all these circles overlap each other has the appearance 
    of two symmetrical ``wings''. The light rays passing at 
    the very edge of the optics result in the ``tail'' ending in a semi-circular contour,
    which delineates the edge of the defocussed image of the optic's pupil. The inset 
    on the left panel shows how the aberration is lined up towards the center of the field, 
    and how the tail's size increases as one moves from center to edge of field.
    Right panel shows an actual star image.
    (graphics by \citealt{coma_graphics}, coma image by \citealt{coma_image}).
}
    \label{fig:coma}
 \end{figure*}

We could in principle use the coma properties delineated in Figure \ref{fig:coma} to tell apart
 images that resulted from light that traversed the telescope, from plate artifacts
 that have no relationship with optics. That way, we could sidestep the problem
 of plate artifacts.

\section{Data and methodology} \label{sec:data}

Data from the {\it Hamburger Sternwarte Doppel-Reflektor} 0.6-m parabolic mirror 
telescope was selected 
for  this project, because its images are affected by significant coma, and because 
enough plate pairs appropriate for the transient search could be found at the archive 
holdings.

A total of 278 plate pairs, assembled from a total of 398 plates, and spanning the years from 1934 to 1957, 
was found to satisfy the Field-Of-View (FOV) overlap and matching exposure time and photographic 
emulsion criteria (Paper I).  To avoid 
contamination by space debris left over by the space programs, which do generate glints 
from reflected Sunlight from mirror-like orbiting debris, October 4, 1957 was 
chosen as sample cutoff because it is the launch date of the first artificial satellite, Sputnik-1.
Most plates used in this work cover FOVs typically about $2.2^{\circ}$ X $1.5^{\circ}$, 
and were scanned at a resolution of 10.583 $\mu\text{m}$/pixel. 
The resulting digital files are typically 12,000 x 8,000 pixels in size.

Table \ref{table:stats} lists some global parameters associated with the data set.

\begin{deluxetable}{lll}
    \tablecaption{Data set parameters\label{table:stats}}
    \tablehead{
        \colhead{parameter}     & 
        \colhead{value}            &
        \colhead{unit}    
    }
    \startdata
    Total plates:                      &         398     & \\
    Total plate pairs:               &         278     & \\
    Total exptime:                   &     121.28    & hour \\
    Total sky area:                  &    1490.09   &  sq. deg \\
    \enddata
\end{deluxetable}

Table \ref{table:table_plates} shows the relevant parameters of the plates where 
transients were found.

\begin{deluxetable*}{ccccc}
    \tablecaption{Plate pairs with transients\label{table:table_plates}}
    \tablehead{
        \colhead{Plate IDs\tablenotemark{a}}   & 
        \colhead{UT 1\tablenotemark{b}}           & 
        \colhead{UT 2\tablenotemark{b}}           &
        \colhead{Exptime (s)} &
        \colhead{Emulsion}
    }
    \startdata
           62525 - 62526 & 1949-04-20 21:56:26  & 1949-04-20 22:51:47  & 1800 & Agfa Astro Z   \\
           62962 - 62963 & 1950-12-09 22:51:38  & 1950-12-09 23:06:35  &  600  & Agfa Astro Panchrom.  \\
           62964 - 62965 & 1950-12-09 23:28:31  & 1950-12-09 23:44:29  &  600  & Agfa Astro Panchrom.  \\
           63037 - 63038 & 1951-02-08 21:09:24  & 1951-02-08 21:29:21  &  600  & Agfa Astro  \\
           63054 - 63055 & 1951-03-04 18:54:08  & 1951-03-04 19:11:06  &  720  & Agfa Astro   \\
           63501 - 63502 & 1953-04-19 22:42:08  & 1953-04-19 23:08:03  & 1200 & Kodak Oa-O   \\
    \enddata
    \tablenotetext{a}{unique IDs generated by APPLAUSE.} 
    \tablenotetext{b}{UT at mid-exposure.} 
    \tablecomments{Plate pairs have the same exposure time and emulsion on both plates.} 
\end{deluxetable*}

The data analysis procedure outlined in Paper I was initially conceived with the 
Schmidt camera images in mind. As such, it partially relies on parameters produced 
by Gaussian fitting to star images, as well as shape information generated by 
computer vision algorithms, in an attempt to filter out artifacts and eventually 
produce a short list of transient candidates. These candidates have in turn to be 
vetted visually, since the filters are far from perfect and let a significant number
of artifacts to be tagged as valid transient candidates. 

For the Doppel-Reflektor data, because of the presence of coma, the processing
was modified in three ways: (i) Gaussian fitting was allowed to work on a more relaxed 
set of bounds;
(ii) filters applied on computer vision shape parameters were turned off, and
(iii) the visual vetting process was speeded up.
The processing procedure also applies filters on a variety of parameters generated by 
SExtractor and made available in the APPLAUSE-generated tables. These filters
were kept in place, since they greatly help to eliminate the most egregious artifacts.

False positives among the remaining candidates were then weeded out by cross-checking
against the USNO catalog \citep{USNO} (the Gaia catalog, \citealt{Gaia_DR3}, is already 
used by the pipeline search algorithm), and visually against the corresponding regions 
in POSS-II blue plates. In some more interesting plates, we actually blinked 
(with {\it ds9}, \citealt{SAOImage}) the entire plate pair in an attempt to 
detect transients that might have eluded the automated procedure.

The visual vetting process used in this work relies on criteria based on the properties
of the coma signature, and how these features actually appear on images within
a range of signal-to-noise levels: (i) comatic images must appear aligned with the direction
that points to the center of the plates; (ii) wings should be visible, and, if not, at least an
asymmetry in the light distribution towards the correct direction of the coma must be visible; 
(iii) the size and visibility of the coma must be consistent with the distance to 
the plate's center and its overall brightness;
(iv) photographic effects such as saturation and halation on bright sources, but especially
possible emulsion reciprocity failure on faint sources, must to be taken into
account when visually evaluating those images.
For each candidate, these features were evaluated against a set of star images 
in the neighborhood of the candidate, with peak flux within 0.1 magnitude of the 
candidate's peak flux.

It should be noted  that the data analysis algorithm described in Paper I
was designed around the concept of {\it vanishing} transients. It cannot detect 
{\it appearing} transients, that is, objects that show up on a plate but do not
appear on the {\it previous} plate.
In some cases, the finding algorithm was run backwards so as to verify the existence of 
these appearing transients. Because this was not done on all existing
plate pairs, it is quite possible that some appearing transients were not detected.

It also should be stressed that these transients were found basically visually. This was
possible because plate artifacts do not show features, at the level examined in this work,
 that can be easily confused with true point source comatic images. Nevertheless, the knowledge 
 gathered with the visual vetting process is fundamental for developing a control sample
 that could be incorporated into a more automated process. 

A full fledged quantitative analysis 
based on measured image parameters will be presented in a forthcoming paper,
where more data from other telescopes will be added in an attempt to address the
shortcomings of the small sample included in this paper. It should be noted that the main
goal of this paper is not to provide results under a statistically rigorous framework, but just
to present evidence that transients are the result of points of light on the sky.

The observed transients are listed in Table \ref{table:table_transients}

\begin{deluxetable*}{rccllrc}
    \tablecaption{Transient properties\label{table:table_transients}}
    \tablehead{
        \colhead{Plate ID\tablenotemark{a}}     & 
        \colhead{Date}          & 
        \colhead{Source ID\tablenotemark{a}} & 
        \colhead{R.A.}           &
        \colhead{Dec.}           &
        \colhead{$m_v$\tablenotemark{b}} &
        \colhead{Annular bin\tablenotemark{c}}
    }
    \startdata
%
           62962 & 1950-12-09 & 40648930008575  &  2 17 54.3   &  +57 37 30  & 10.9  &  2  \\
           62964 & 1950-12-09 & 40648950008626  &  2 24 06.1   &  +57 22 13  & 10.6  &  6   \\
           63037 & 1951-02-08 & 40649680004453  &  2 17 24.6   &  +57 33 08  & 10.0  &  2  \\
           63038 & 1951-02-08 & 40649690003944  &  2 14 49.7   &  +57 33 57  & 10.6  &  4  \\
           63054 & 1951-03-04 & 40649850007759  &  2 28 28.3   &  +57 07 55  & 7.1    &  6   \\
           63054 & 1951-03-04 & 40649850009785  &  2 19 00.5   &  +57 31 53  & 7.9    &  3  \\
           63054 & 1951-03-04 & 40649850006760  &  2 16 49.5   &  +57 06 38  & 9.1    &  4  \\
           62525 & 1949-04-20 & 40644250018738  &  13 13 37.4 &  +17 13 43  & 11.9  &  4    \\
           63501 & 1953-04-19 & 40654930007682  &  13 14 07.1 &  +18 08 31  & 10.2  &  1   \\
           63501 & 1953-04-19 & 40654930007722  &  13 14 08.8 &  +18 08 07  & 10.4  &  1   \\
           63501 & 1953-04-19 & 40654930007709  &  13 14 08.0 &  +18 07 47  & 10.5  &  1   \\
    \enddata
    \tablenotetext{a}{unique IDs generated by APPLAUSE.} 
    \tablenotetext{b}{V magnitude derived from APPLAUSE-computed data.} 
    \tablenotetext{c}{Distance to plate center: 1 - center, 9 - edge} 
\end{deluxetable*}

\section{Results} \label{sec:results}

\begin{figure*}[ht!]
\plotone{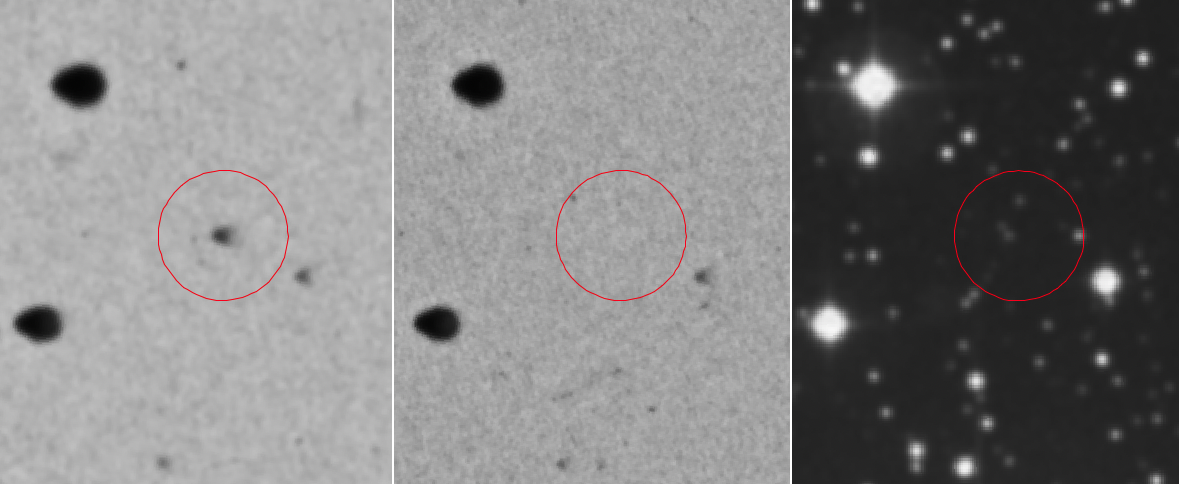}
\caption{Transient 40649850006760 from plate pair 63054 (left) and 63055 (center). 
Negative (photographic density) scale is used to highlight faint detail. 
See Figure \ref{fig:transient_7a} for a detail of the transient itself. 
Right panel depicts the same field as it appears on the POSS-II blue plate.}
\label{fig:transient_1}
\end{figure*}

\begin{figure*}[ht!]
\plotone{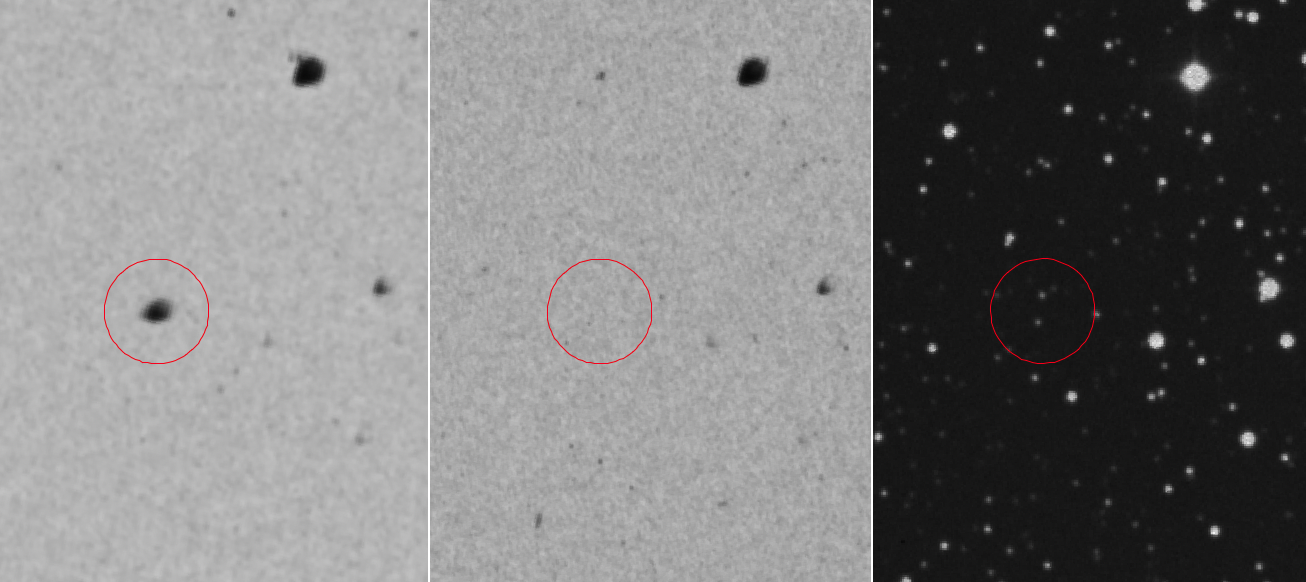}
\caption{Transient 40649850009785 from plate pair 63054 (left) and 63055 (center). 
Negative (photographic density) scale is used to highlight faint detail. 
See Figure \ref{fig:transient_7a} for a detail of the transient itself. 
Right panel depicts the same field as it appears on the POSS-II blue plate.}
\label{fig:transient_2}
\end{figure*}

\begin{figure*}[ht!]
\plotone{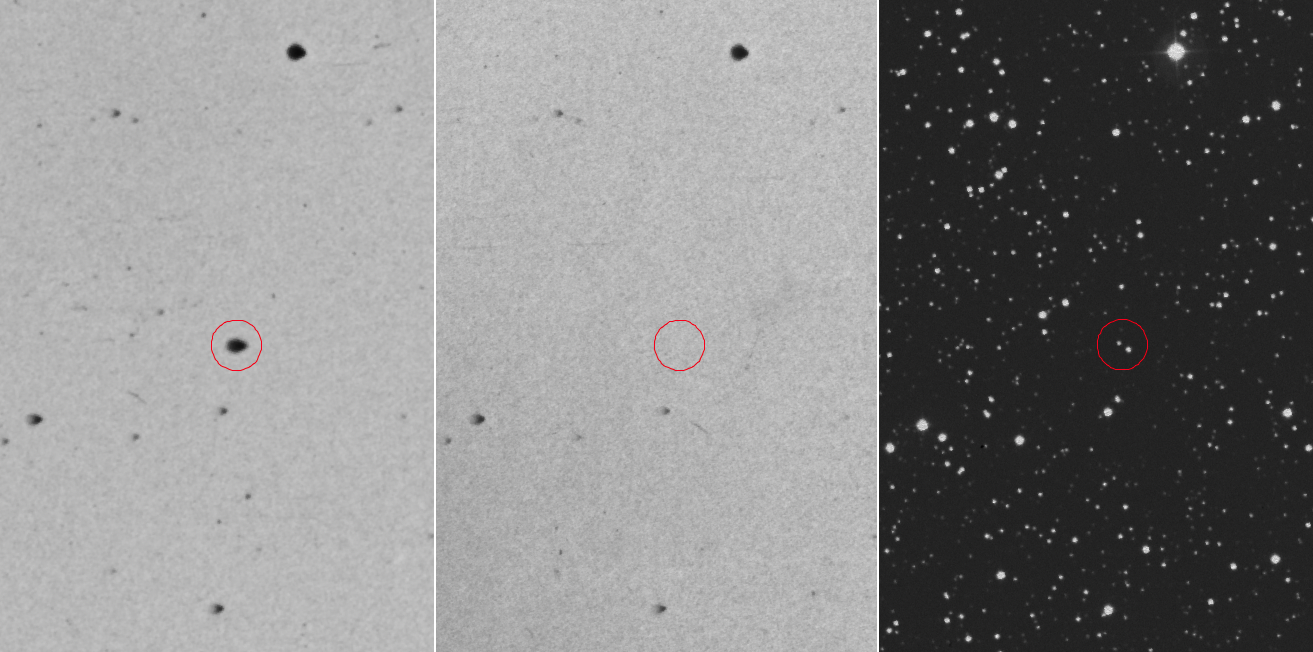}
\caption{Transient 40649850007759 from plate pair 63054 (left) and 63055 (center). 
Negative (photographic density) scale is used to highlight faint detail. 
A larger, 15\arcmin X 8\arcmin field of view is show in order to display the neighborhood of the event. 
See Figure \ref{fig:transient_7a} for a detail of the transient itself. 
Right panel depicts the same field as it appears on the POSS-II blue plate.}
\label{fig:transient_3}
\end{figure*}

\begin{figure*}[ht!]
\plotone{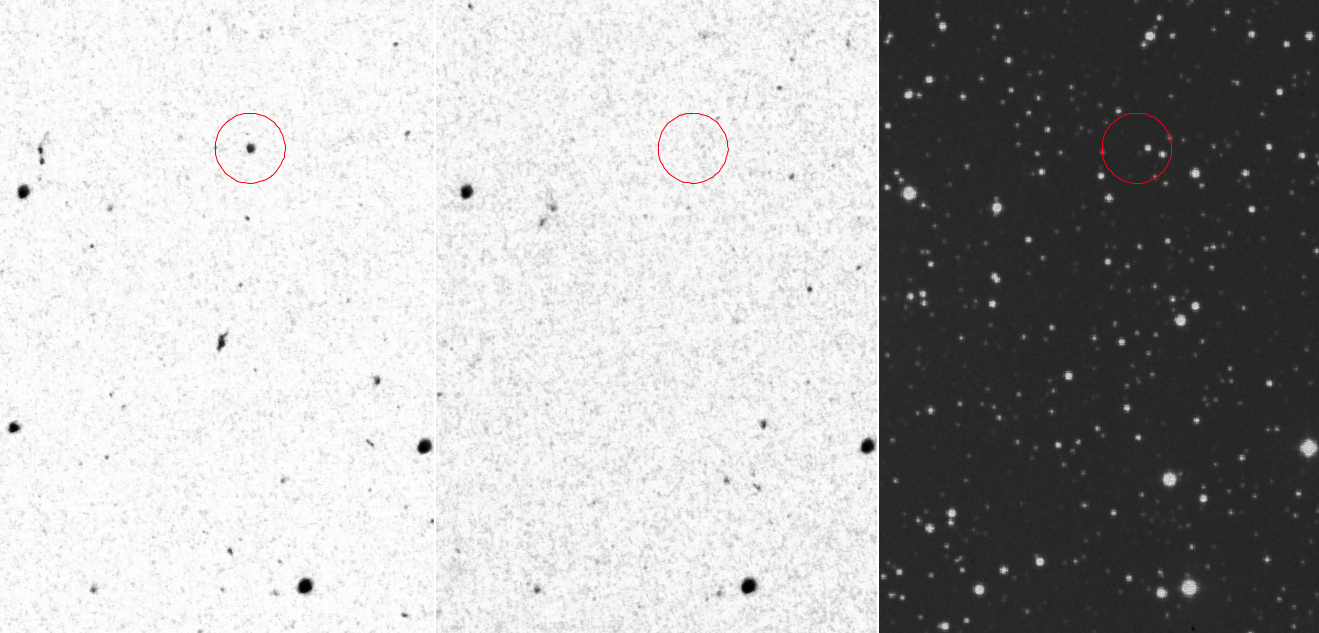}
\caption{Transient 40649680004453 from plate pair 63037 (left) and 63038 (center). 
Negative (photographic density) scale is used to highlight faint detail. 
See Figure \ref{fig:transient_7a} for a detail of the transient itself. 
Right panel depicts the same field as it appears on the POSS-II blue plate.}
\label{fig:transient_6}
\end{figure*}

\begin{figure*}[ht!]
\plotone{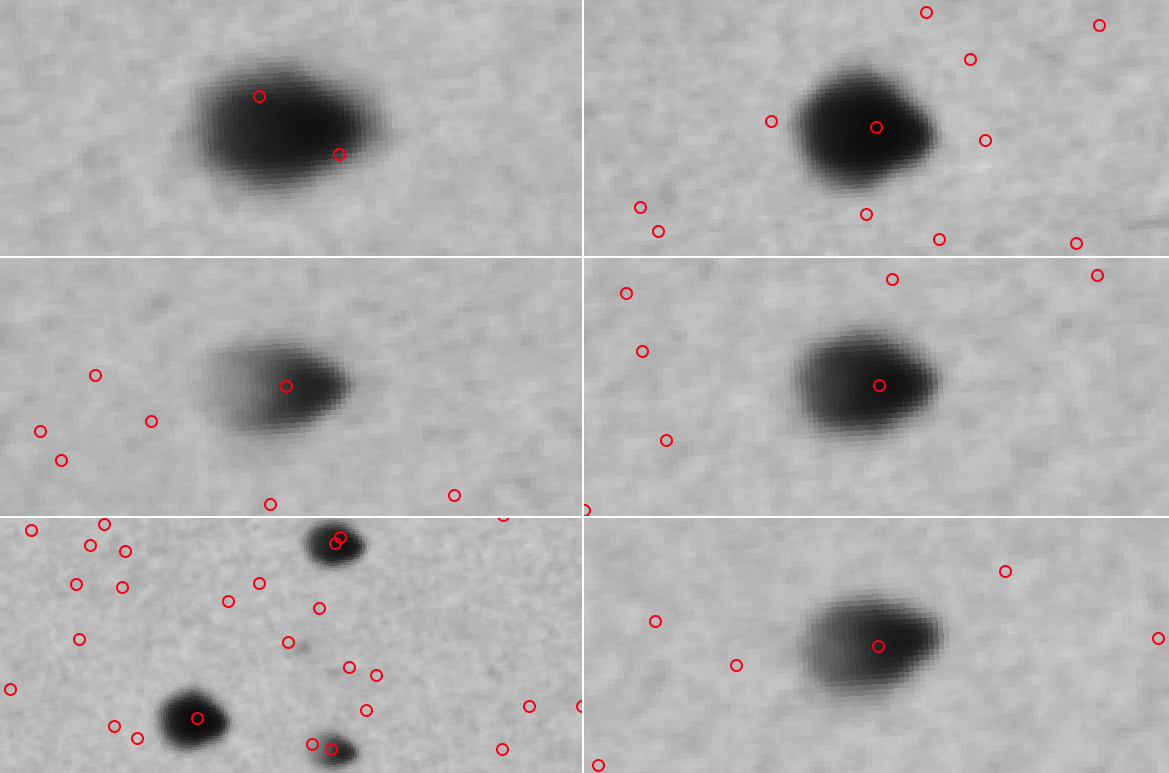}
\caption{Detail of transient 40649850007759 at top left shows the position of USNO 
reference stars near its location. The other panels depict some typical stars near the 
transient location. They illustrate how their USNO catalog position relates to the 
comatic star image. That position is typically near the apex of the triangular 
wing structure, symmetrically centered on it. It is the point of peak intensity \citep{coma}.
This is observed in all stars on that particular plate. On the transient image, on the 
other hand, we do not see that, but instead we see the off-centered positions 
of the two faint stars visible within the circle in the POSS-II image in 
Figure \ref{fig:transient_3}. 
This shows that the origin of the transient is not likely to be one of those two stars. 
Negative (photographic density) scale is used to better highlight faint detail and 
structure of the coma images.}
\label{fig:transient_3a}
\end{figure*}

\begin{figure*}[ht!]
\plotone{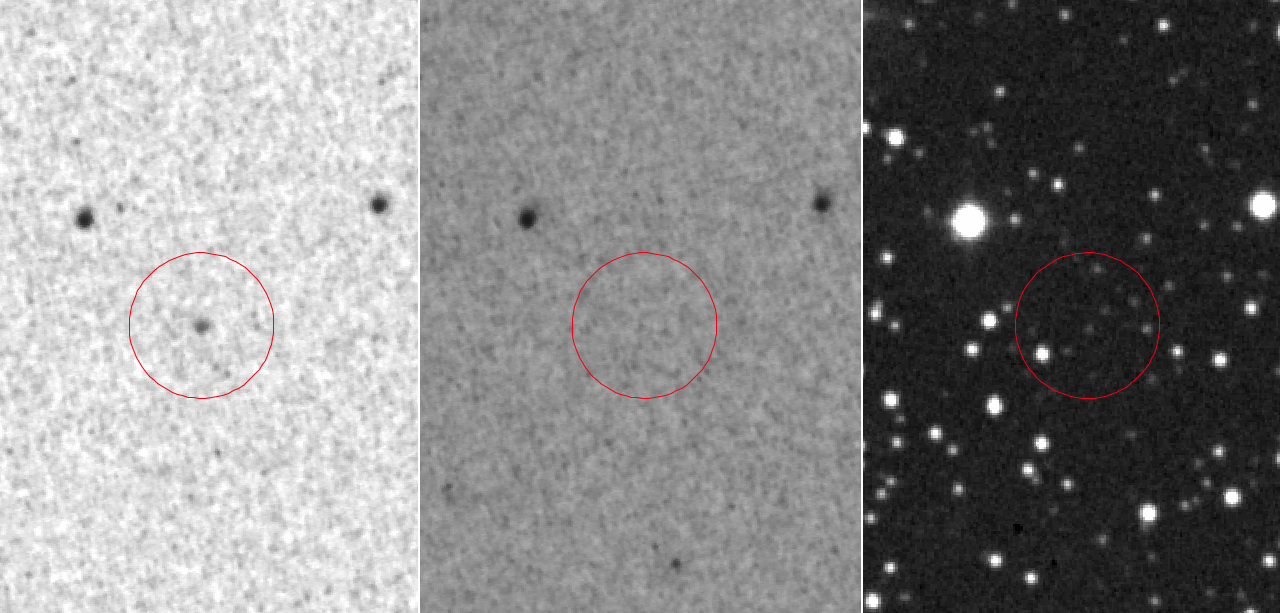}
\caption{Transient 40648930008575 from plate pair 62962 (left) and 62963 (center). 
Negative (photographic density) scale is used to highlight faint detail. 
See Figure \ref{fig:transient_7a} for details of the transient itself. 
Right panel depicts the same field as it appears on the POSS-II blue plate.}
\label{fig:transient_9}
\end{figure*}

\begin{figure*}[ht!]
\plotone{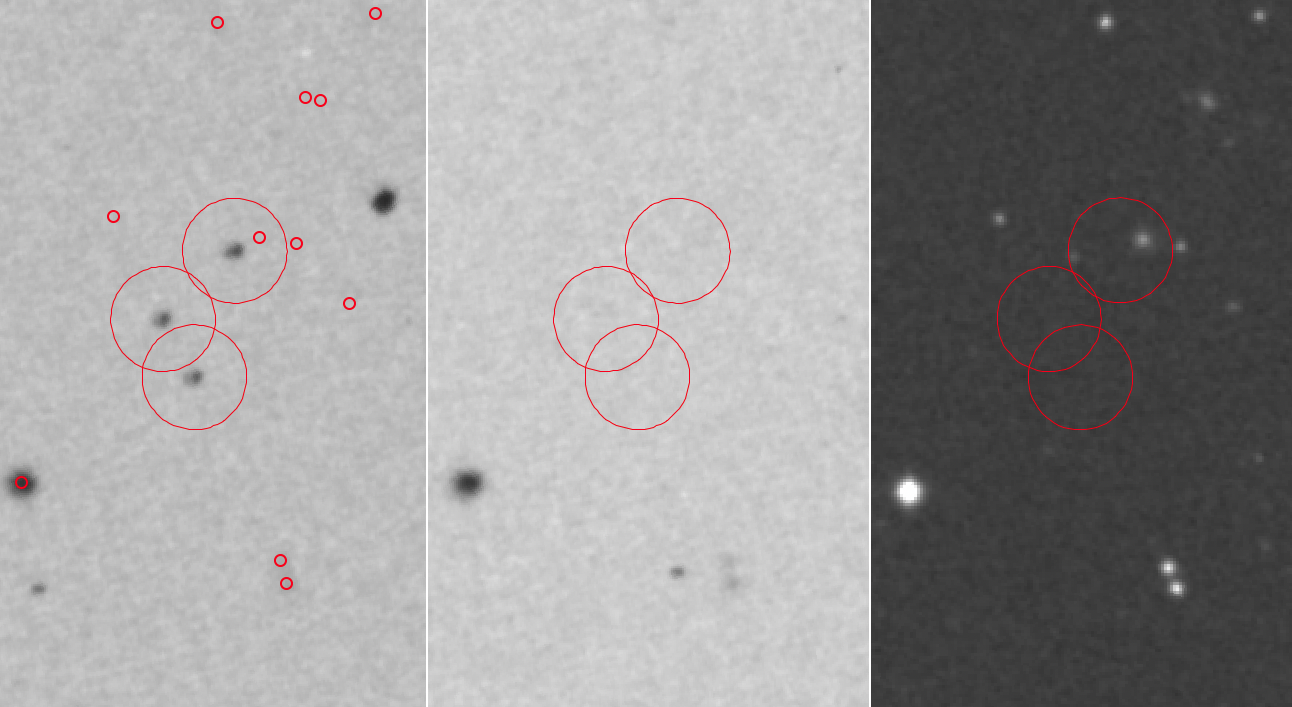}
\caption{Three transients from plate pair 63501 (left) and 63502 (center). 
Negative (photographic density) scale is used to highlight faint detail. 
Small red circles indicate star positions from the USNO catalog.
See Figure \ref{fig:transient_7a} for details of the transients themselves. 
Right panel depicts the same field as it appears on the POSS-II blue plate.}
\label{fig:transient_7}
\end{figure*}

\begin{figure*}[p]
\centering
\includegraphics[width=\textwidth,height=\textheight,keepaspectratio]{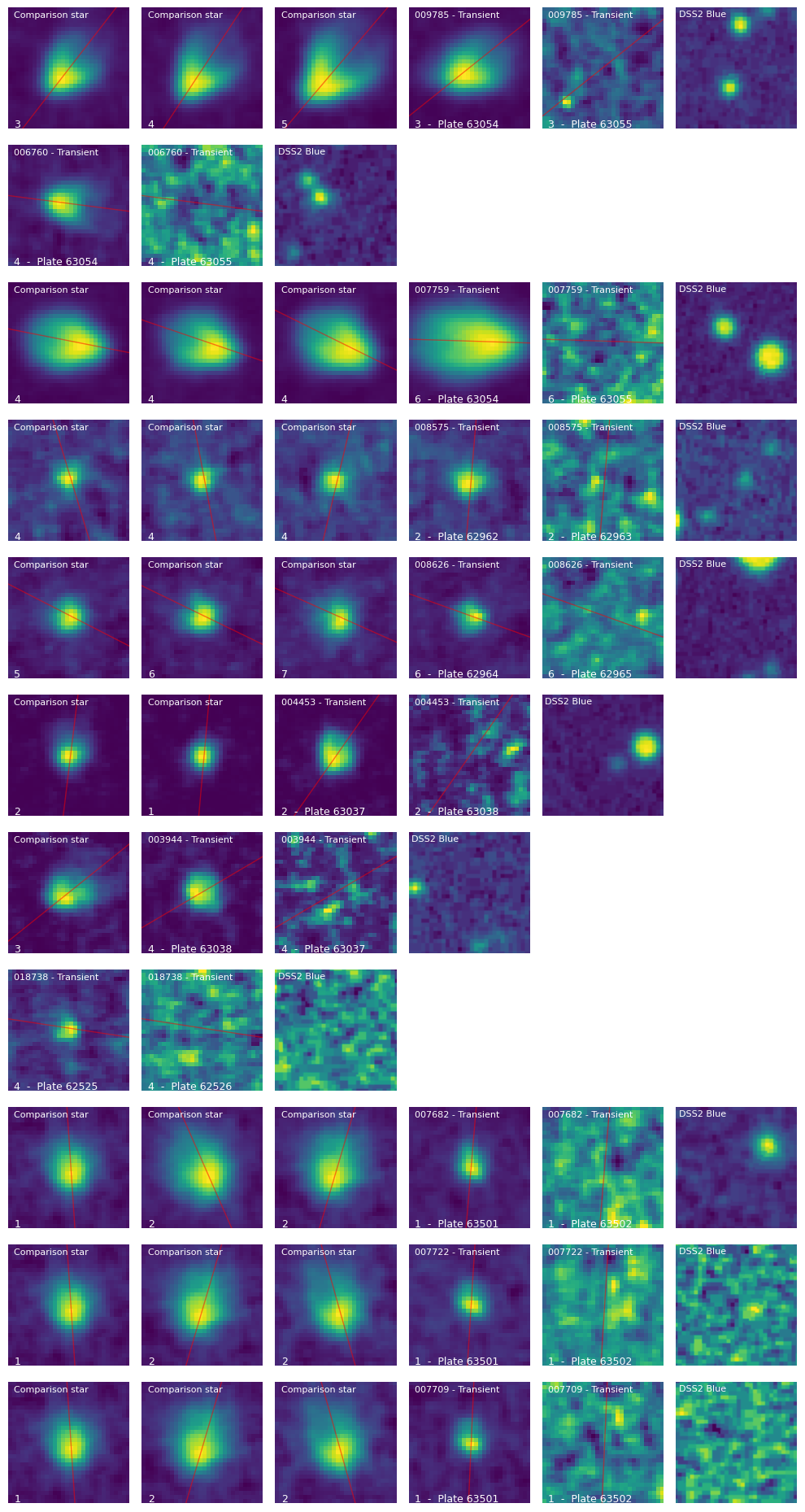}
\caption{Transients displayed together with comparison stars in their respective neighborhoods. 
Comparison stars were chosen within 0.1 magnitude of the transient's peak flux.
The transient is identified by the last 6 digits of its source ID (as per Table \ref{table:table_transients}). 
The numbers at the bottom indicate the {\it annular bin}
(APPLAUSE concept to codify distance to center of plate: 1 - center, 9 - edge).
The transient is displayed as it appears in both images of the corresponding plate pair.
The rightmost image is how the region (with same size and scale) appears on the POSS-II blue plate.
The red line indicates the direction of the coma, as expected from optical geometry alone.
Pseudo-color log photographic density scale is used to balance visibility in between the faint 
structures at the wings, and the bright core.}
\label{fig:transient_7a}
\end{figure*}

Images of transients are shown in Figures \ref{fig:transient_1} to \ref{fig:transient_7}.
Detailed close-up images of all transients listed in Table \ref{table:table_transients} are shown
together in Figure \ref{fig:transient_7a}
These images are direct evidence that the transients resulted from
light that actually traversed the optical train of the telescope, thus invalidating
the plate artifact argument, {\it at least for these transients}. It does not mean that other
transients already reported in the literature were not originated by plate artifacts.  

Of note is the fact that transients cover a relatively wide range of brightnesses, including three 
very bright ones, all on the same plate on March 4, 1951. Nothing can be said about
the other end of the brightness scale, however, since our sample is incomplete in a
variety of senses. The weakest transient identified as such would suggest that the
sample would have a limit of around  $m_v \approx 11.$ Concerns about confirmation
bias on a basically visual identification method are valid at these lower brightnesses,
but the sample also includes brighter transients that find no counterpart on known
plate artifacts that superficially have the appearance of comatic images of point sources. 

One should note that we do not know anything about the actual duration of each transient. The 
magnitudes in Table \ref{table:table_transients} assume that the transient behaved
like a star, that is, was exposed by the same exposure time everything else on the
plate was. But in case they are events of shorter duration than the plate exposure time, 
the actual
brightness would be proportionaly larger. For example, a $m_v \approx 10.$ transient that
lasted 1 sec. and was recorded on a plate with 15 min. exposure time, would
have an actual brightness of $m_v \approx 2.6$.

The detected events appear clustered in both time and sky position; however, 
the statistical significance of this pattern cannot yet be assessed because of the 
non-uniform sampling of the dataset.
Table \ref{table:table_transients} shows
seven transients detected on a patch of sky around RA=2:17, Dec=+57, and
four transients similarly detected close on the sky, around  RA=13:14, Dec=+18.
This all happens in a window between 1949 and 1953, other years being devoid
of detections. Note that the time sampling is very non-uniform, with about only 20 
plate pairs in between 1934 and 1949, and about 70 plate pairs in between 1953 
and 1957, the sample cutoff. 

\begin{figure*}[ht!]
\plotone{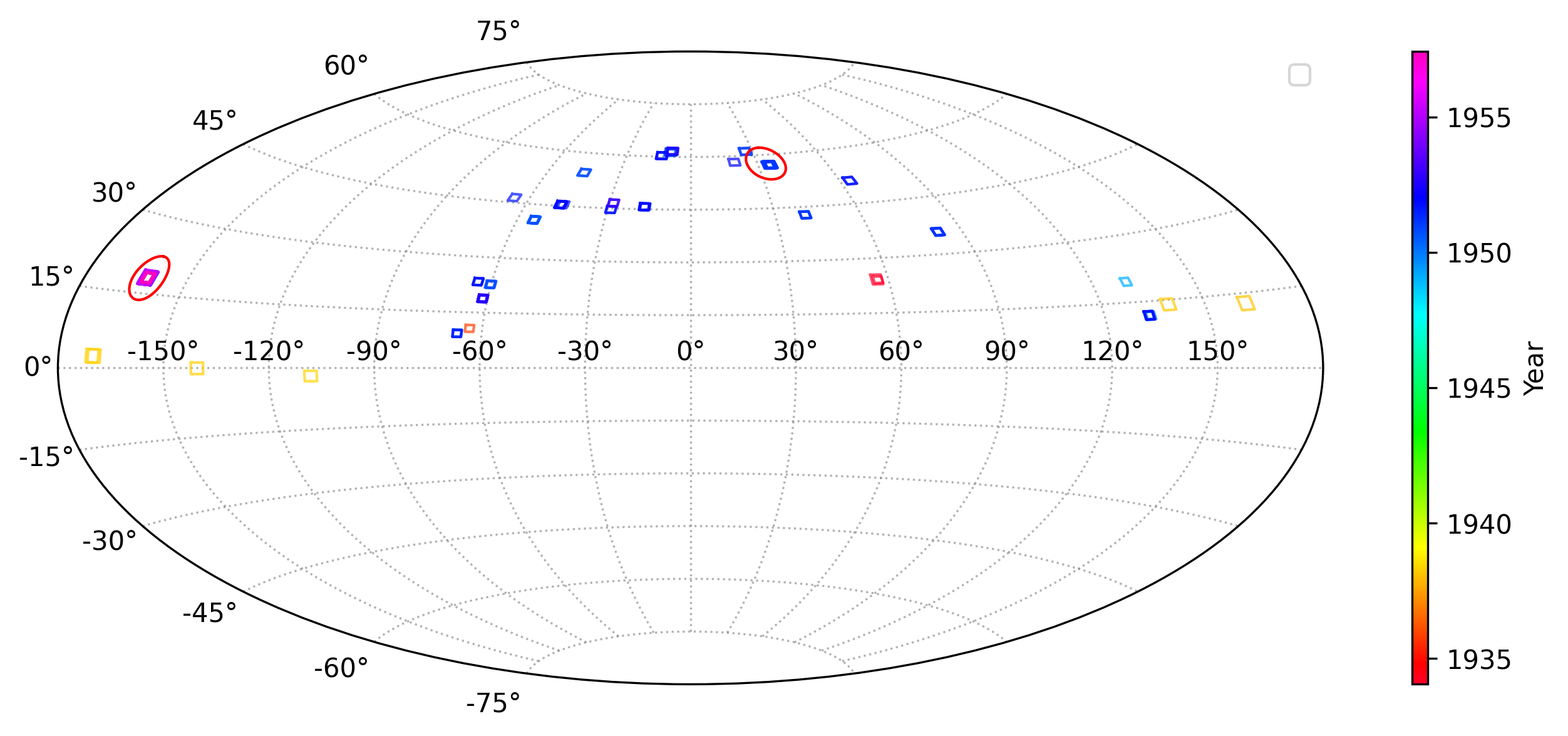}
\caption{Spatial distribution of the entire data set used in this work. Each square represents
a plate pair. 
The plate dates are color-coded; the darkness of each frame act as a rough measure of
the number of plates taken at a given position on the sky. The two circled regions are
the ones where transients were found.}
\label{fig:aitoff_plot}
\end{figure*}

\begin{figure*}[ht!]
\plotone{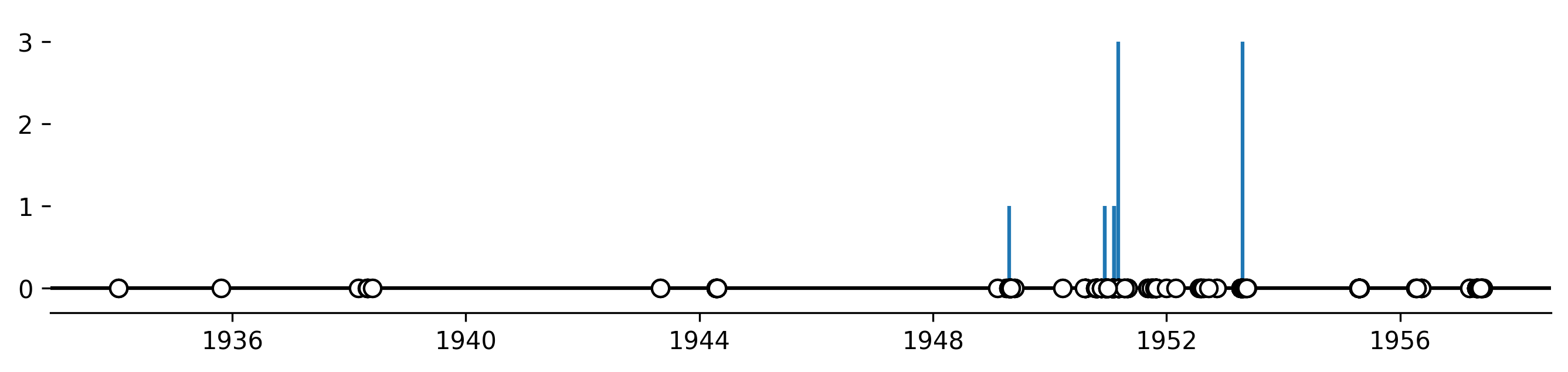}
\caption{Temporal distribution of the entire data set used in this work. Each circle at the
baseline represents one plate pair. Number of detected transients in each plate pair is
depicted by the vertical bar.}
\label{fig:timeline_plot}
\end{figure*}

\begin{figure*}[ht!]
\plotone{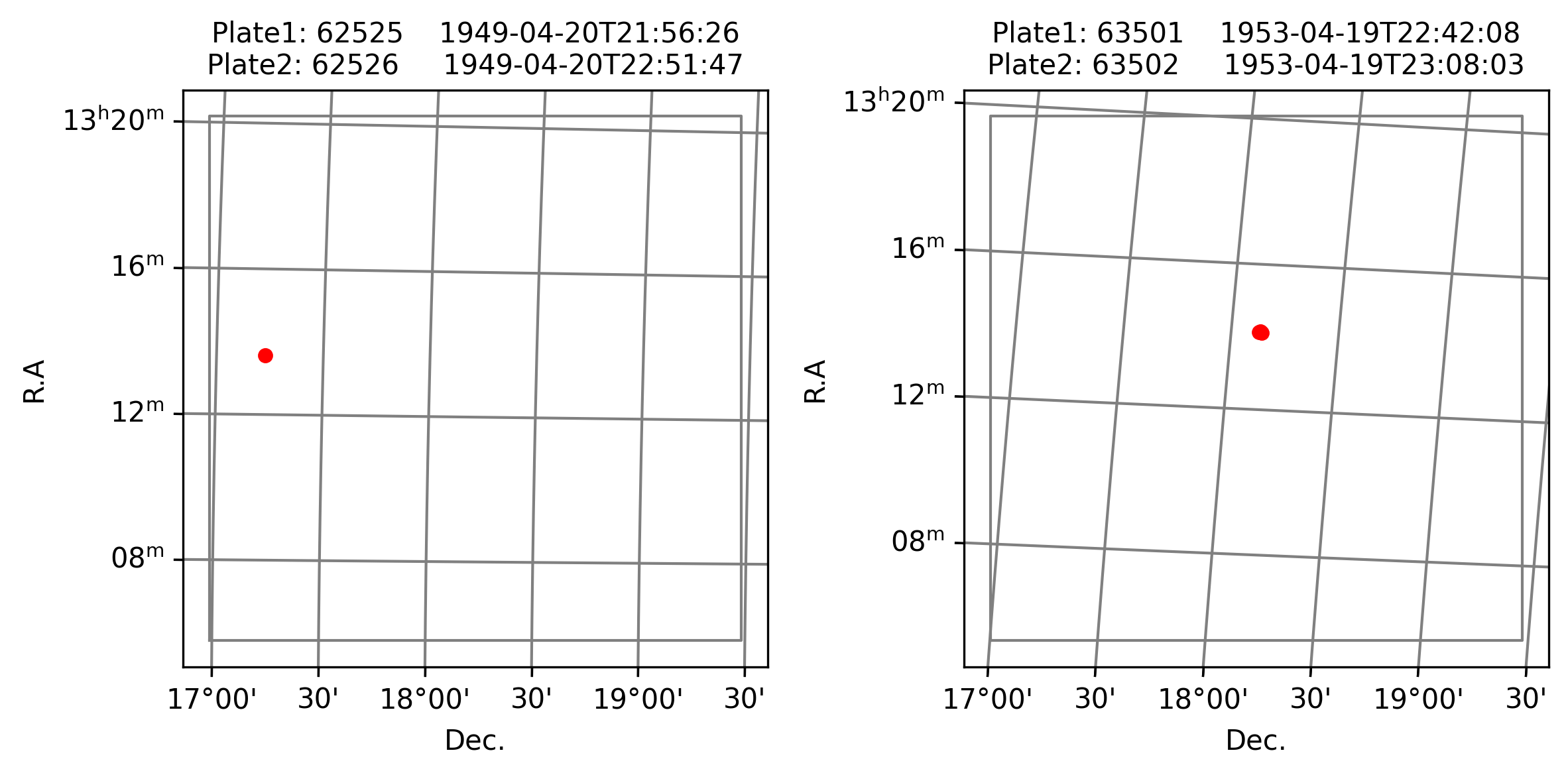}
\caption{Transients that appear in the same field, at exactly four years apart from each other.
Note that the right panel displays not one, but three transients very close together, as in 
Figure \ref{fig:transient_7}}
\label{fig:frame1_plot}
\end{figure*}

\begin{figure*}[ht!]
\plotone{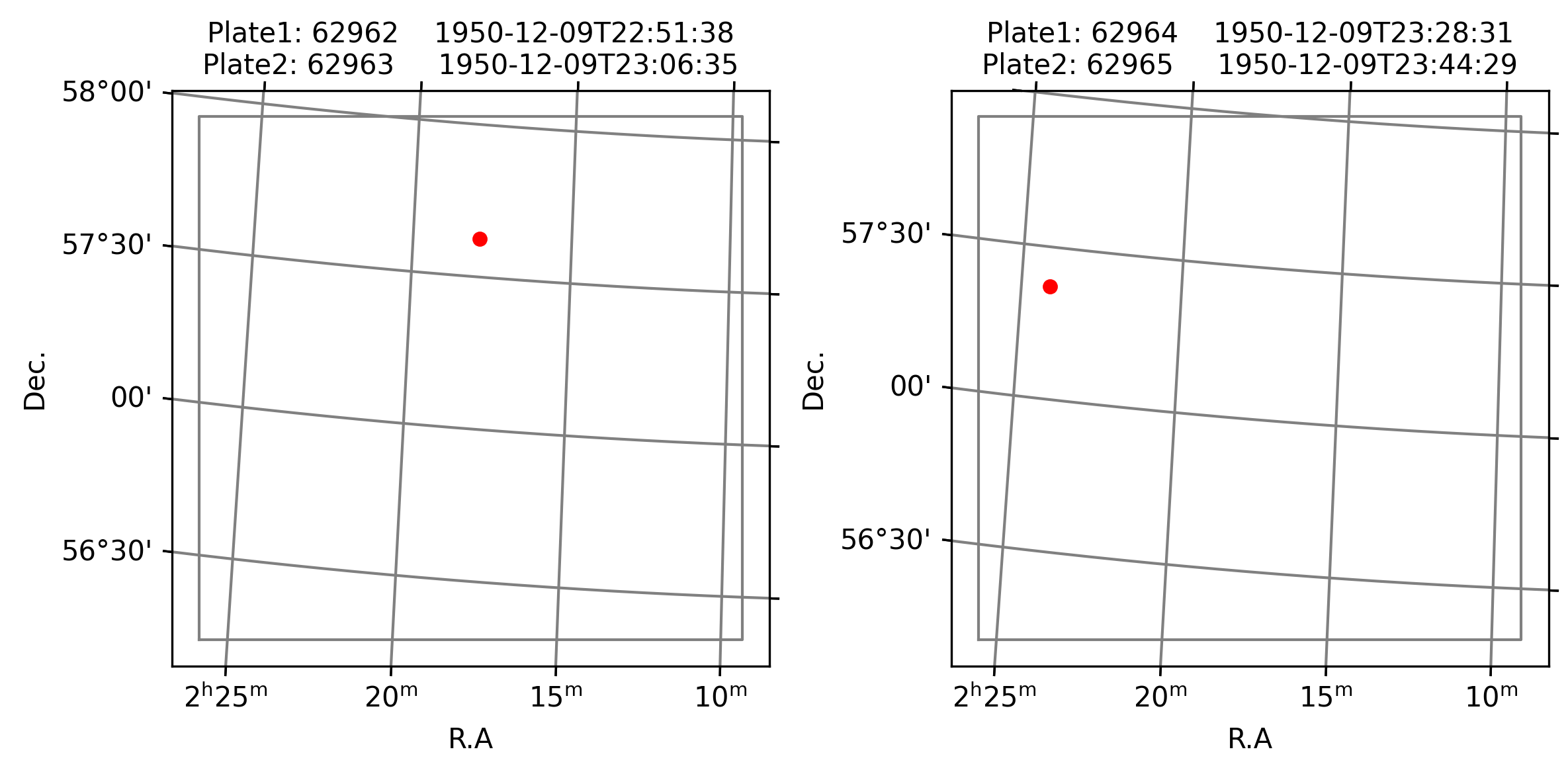}
\caption{Two transients that appear in the same field, right after each other. One could speculate
that they were caused by a single moving source.}
\label{fig:frame2_plot}
\end{figure*}

\begin{figure*}[ht!]
\plotone{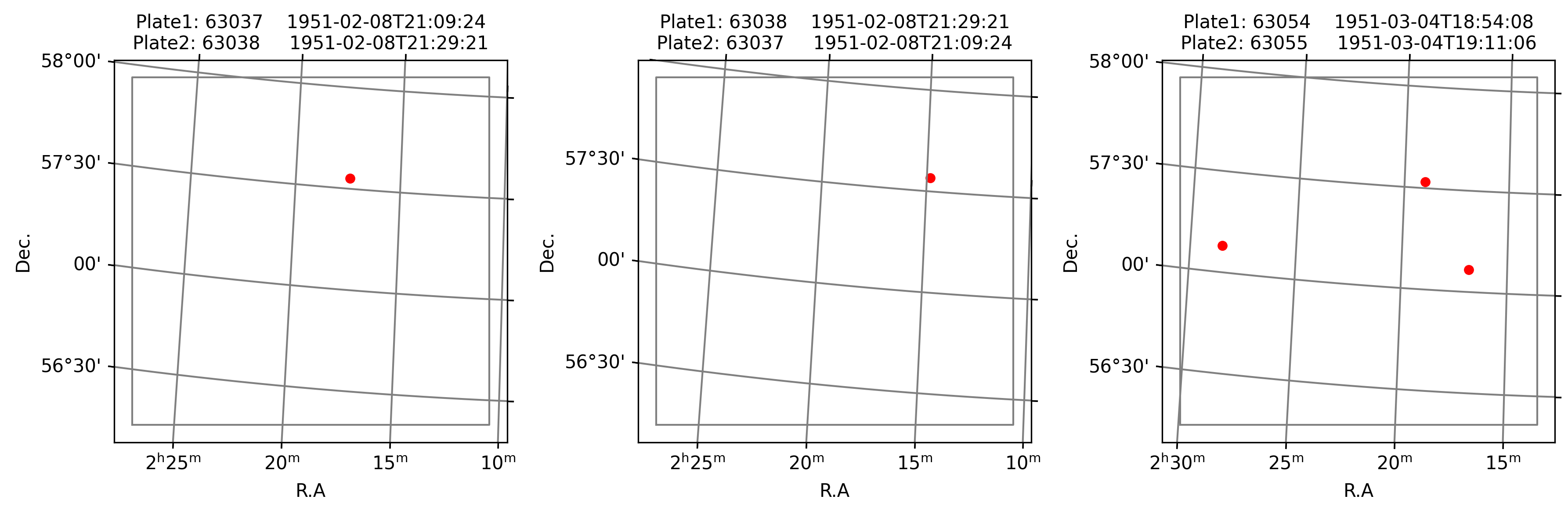}
\caption{Transients in the same field. Note that the second panel depicts an \textit{appearing} transient,
since the pipeline was run backwards on the plate's dates.}
\label{fig:frame3_plot}
\end{figure*}

Figures \ref{fig:aitoff_plot}, \ref{fig:timeline_plot}, \ref{fig:frame1_plot}, \ref{fig:frame2_plot},
and \ref{fig:frame3_plot} attempt 
to summarize the spatial and temporal sampling of the entire data set.

On December 9, 1950, a vanishing transient on plate 62962 is followed by 
another vanishing transient on plate 62964. Their mid-exposure times are 
separated by 30 minutes, and they are about $1.5^{\circ}$ apart on the sky
(Figures \ref{fig:transient_9} and \ref{fig:frame2_plot}).

On the same vein, on February 8, 1951, a similar sequence of events shows 
up {\it in the same sky field as the December 9, 1950 transients}: 
a vanishing transient on plate 63037, followed by an
appearing transient on the plate immediately after, 63038. Their mid-exposure
times are separated by 20 minutes, and they are separated on the sky by about
35\arcmin on an almost E-W direction. Their shapes and brightnesses look similar,
since they are close on the plate, thus have similar coma structure 
(Figures  \ref{fig:transient_6} and \ref{fig:frame3_plot}). 

Asteroids can be ruled out as the cause of these transients, however, since no 
blurring is visible on the 10 min. long exposures, and the distance travelled is 
inconsistent with typical asteroid speeds. 
The Minor Planet Center asteroid finding tool also returns empty for both
date, time, coordinates data sets. 

Following the above, on March 4, 1951, three very bright events were recorded again
on the same patch of sky as the ones described above 
(Figures  \ref{fig:transient_1}, \ref{fig:transient_2}, \ref{fig:transient_3}, \ref{fig:transient_3a},
and \ref{fig:frame3_plot}). Of note is the fact that the photographic image of the
brightest event shows evidence of {\it saturation} of the emulsion response to light,
and {\it halation}, the result of light backscatter in the emulsion layer. These further
confirm that the image resulted from light acting on the emulsion, and not a plate defect. 

The other four transients are also remarkable, because one happened on
April 20, 1949, but the other three, on April 19, 1953, showed up {\it in the same
sky field as the April 20, 1949 transient}, all three packed
together in a very small region of about 45\arcsec
(Figures \ref{fig:transient_7}, \ref{fig:transient_7a}, and \ref{fig:frame1_plot}). One could
argue against the classification of these three events as true transients,
given their appearance, significantly narrower than stars of same peak
flux. On the other hand, that might be in fact a clue that they were produced
by brief pulses of light. Because the temporal power spectrum of
atmospheric-induced seeing falls very quickly with frequency 
($\propto f^{-8/3}$, \citealt{tatarskii1971}),
most of what the photographic plate records as seeing is caused by image
wander. A short flash of light will be less affected by that effect, and thus
will result in a sharper image. However, that image will still carry the coma 
signature. That signature might in turn be rendered less visible, in weak images,
 due to emulsion reciprocity failure at these short exposure times. 
 The appearance of the three recorded  events can be explained in those grounds
 (other weak transients in this work partially share the same properties). 
 It seems rather unlikely that all three 
 would be caused by plate artifacts that happen to have an asymmetry of same 
 magnitude and, most importantly, oriented in the same direction exhibited 
 by the coma on neighboring star images.
 It should be noted that the concept of transients looking sharper than stars 
 due to short exposure times was already proposed by \cite{V2025b}.

Given the patchy sampling of sky and timeline imposed by the data set 
used in this work, one cannot rule out that some, or all, those spatial and temporal 
groupings described above are just caused by sampling. Although that would 
imply that these transients are commonplace, and wherever and whenever
one looks, one will be bound to easily find them. The groupings might as well be
interpreted as the effect that a single, moving object, could be the cause of
multiple grouped transients, although we do not have enough data to reach 
any firm conclusion in that regard. These topics will be better addressed on a 
forthcoming paper were more data from other telescopes will be added in 
order to increase the sample size to a point were meaningful statistical 
inferences can be made. 

  Of note are also possible associations present in our sample, with nuclear weapon testing. 
These associations were first observed by \citealt{Bruehl2025} in the POSS data.
Several events in the presented sample occurred within days or weeks of atmospheric 
nuclear tests. Whether this reflects a genuine correlation, or a chance coincidence, cannot 
be assessed with the present sample.

The following presents some possible associations that might exist in our data set.

\begin{itemize}
    \item two weak transients happened two days after Operation Ranger, conducted 
    in the Nevada Proving Ground between January 27 and February 6, 1951,
    executed the last of five explosions in the test run. 
    \item three transients, the brightest in our sample, happened on March 4, 1951, 
    about one month after the last explosion in the Ranger test run. 
    \item a triple transient occurred Apr 19, 1953, one day after Operation 
    Upshot-Knothole run the 6th explosion in their test run, after other five that 
    started on March 17.
\end{itemize}

Then again, in 1950 there were no nuclear tests in the entire world, and our sample
has two transients in that year. In Aug 29 1949 the USSR conducted the only 
nuclear test that year, in the world, and our only transient in that year predates the test
by about four months.

\section{Conclusions} \label{sec:conclusions}

Results presented in this paper provide independent evidence for the reality of 
fast transients in archival
photographic plates, using a data set and methodology independent of other previous
work on the subject performed by the VASCO Project. Although that work produced
solid evidence of the reality of the transient phenomenon, there is still some lingering 
skepticism about it, mostly
based on the argument that such transients can be mimicked by certain photographic
plate artifacts. 

The search presented here, performed on data made available by the APPLAUSE 
Archive, yielded images of transients that contain the optical signature of telescope coma, which
is strong evidence that these  were created  
by light that passed throughout the optics of the telescope, and not by plate artifacts.

The data set consists of 398 plates produced by the Hamburger Sternwarte 
Doppel-Reflektor 0.6-m telescope, during the period 1934 to 1957. That telescope 
creates images that are mildly affected by coma. Using a slightly modified methodology 
as the one presented in Paper I, we found eleven transients on that data set.

The transients exhibit a remarkable degree of clustering, both in time as
in space. All eleven showed up on only two small regions of the sky, even when 
appearing on separate nights.
They also all happened in the period 1949 - 1953, but that might be a sampling
effect, since about two thirds of the entire set of plates were taken in that time
interval.

Astronomical hypotheses for the origins of transients were examined by 
\citealt{Solano_2022_MNRAS}; none had success in satisfactorily explaining 
their know properties.
Non-astronomical explanations were also advanced by several authors.
Hypotheses based on plate artifacts \citep{Hambly2024} are not supported by the 
findings in this work, although we still need to further develop the comatic-image technique 
with better sampling and an automated identification algorithm.
\citealt{Villarroel_2021} shows that micrometeorites seen face-on are not a good explanation. 
Ghosting and internal reflections on the optics can be dismissed by the fact the present work
relies on pairs of plates: if ghosting from star images in the field happened, it should show
up on both plates and thus be automatically excluded from the results.
Hypotheses based on little-known upper atmosphere effects still need to be further
developed.

However, the data presented here are consistent with two non-astronomical 
hypotheses advanced in recent years. 
Although the plate 
sample is not complete and homogeneous enough for us to draw a statistically
meaningful  conclusion, the data is consistent with the association that \cite{Bruehl2025} 
found to exist  between transients and nuclear weapon testing. 
On the same token, the data are also consistent with the hypothesis that these 
transients may be Sunlight glints generated by tumbling, mirror-like objects in 
space, in the vicinity of Earth \citep{villaroel2025aligned}. 

\section*{Acknowledgments}

Funding for APPLAUSE has been provided by DFG (German Research Foundation, Grants
EN926/3-1/2, HE1356/63-1/2, GR969/4-1, SCHM1032/552), 
Leibniz Institute for Astrophysics Potsdam (AIP), Dr. Remeis Sternwarte Bamberg (University 
Nurenberg/Erlangen), the Hamburger Sternwarte (University of Hamburg) and Tartu Observatory.

This research has made use of data and/or services provided by the International 
Astronomical Union's Minor Planet Center
\footnote{\url{https://www.minorplanetcenter.net/cgi-bin/checkmp.cgi}}

The author wishes to acknowledge the valuable suggestions provided by two anonymous
reviewers.

\subsection*{Data sources}

Exposure data for all plate pairs used in this work, in the form of a FITS table, can be downloaded from here:
\url{https://github.com/cuernodegazpacho/plateanalysis/blob/main/footprints/results/DR/exposure_data_DR.fits}

The table contents were taken straight from the APPLAUSE database.

\subsection*{Software}
plateanalysis \citep{ivo_busko_2026},
astropy \citep{astropy:2013, astropy:2018, astropy:2022},
Source Extractor \citep{1996A&AS..117..393B},
numpy \citep{harris2020array},
{\it ds9}, \citep{SAOImage}

\bibliography{main}{}
\bibliographystyle{aasjournalv7}

\end{document}